\documentclass{aa}  
\usepackage{latexsym, epsfig}
\usepackage{graphicx}
\usepackage{natbib}
\usepackage{txfonts}

\def\kms{kms$^{-1}$~}

\begin{document}
   \title{Spatially resolved H$_2$ emission from the disk around T Tau
     N\thanks{Based on observations collected at the European Southern
       Observatory, Paranal, Chile under the programme 60.A-9041(A)}}


   \author{M. Gustafsson
          \inst{1}          \and
          L. Labadie \inst{1} \and T. M. Herbst \inst{1} \and M. Kasper \inst{2}}
          
   \offprints{M. Gustafsson}

   \institute{Max Planck Institute for Astronomy, K\"
{o}nigstuhl 17, 69117 Heidelberg, Germany \\
              \email{gustafsson@mpia.de, labadie@mpia.de, herbst@mpia.de}
    \and European Southern Observatory, Karl-Schwarzschild-Str. 2, 85748
    Garching, Germany, \email{mkasper@eso.org}}

   \date{Received 12 March 2008; accepted 18 June 2008}

 
  \abstract
   { Molecular hydrogen is the main constituent of circumstellar disks and
     could be an important tracer for the evolution and structure of such
     disks. So far, H$_2$ has only been detected in a few
     disks and only through spectroscopic observations, resulting in
     a limited 
     knowledge of the spatial distribution of the H$_2$ emitting gas. 
   }
   {We report the detection of quiescent H$_2$ emission
     in a spatially resolved ring-like structure within 100
     AU of T Tau N. We present evidence to show that the
     emission most likely arises from shocks in the atmosphere of a nearly
     face-on disk around T Tau N.}
   {Using high spatial resolution 3D spectroscopic K-band data, we trace
     the spatial distribution of several H$_2$ NIR rovibrational lines in the
     vicinity 
   of T Tau N. We examine the structure of the circumstellar material around
   the star through SED modeling. Then, we use models of shocks and UV+X-ray
   irradiation to 
  reproduce the H$_2$ line flux and line ratios in order to test how the H$_2$
  is excited.}
   {We detect weak H$_2$ emission from the v=1-0 S(0), S(1), Q(1) lines and 
     the v=2-1 S(1) line in a ring-like structure around T Tau N
     between 0\farcs1 ($\sim15$AU) and 0\farcs7 ($\sim$100AU) from the
     star. The v=1-0 S(0) and v=2-1 S(1) lines are detected only in the outer
     parts 
     of the ring structure. Closer to the star, the strong continuum limits our
     sensitivity to these lines. The total flux of the v=1-0 S(1) line is
     $1.8\times10^{-14}$ergs s$^{-1}$cm$^{-2}$, similar to previous
     measurements of H$_2$ in circumstellar disks.     
The velocity of the H$_2$ emitting gas around T
     Tau N is consistent with the rest velocity of the star, and the H$_2$ does
     not 
     seem to be part of a collimated outflow. Both shocks impinging on the
     surface of a disk and 
     irradiation of a disk by
     UV-photons and X-rays from the central star are plausible candidates for
     the H$_2$ excitation mechanism. However, irradiation should not create
     a large degree of excitation at radii larger than 20 AU.
      Most likely the H$_2$ emission arises in the
     atmosphere of a flared disk with radius 85-100 AU and mass
    0.005-0.5M$_{\odot}$, where the gas is excited by shocks created when a
     wide-angle wind impinges on the disk. The H$_2$ emission could
    also originate from shock excitation in the cavity walls of an envelope,
    but this requires an unusually high velocity of the wide-angle wind from T
    Tau N. 
   } 
   {}

   \keywords{Stars: winds, outflows, circumstellar matter, Stars:
     emission-line, Stars: pre-main sequence,  Infrared: stars  }
   \maketitle
%

\section{Introduction}

The study of circumstellar disks around young stars is essential to
understanding their evolution from gaseous disks to planetary systems. 
In this paper, we examine the spatial distribution of molecular hydrogen, the
main constituent of disks around young stars. 
Disks have been observed in a wide range of wavelengths ranging from
optical to millimeter, although only a few studies have concentrated on the
H$_2$ component. Many investigations have focused on the broad band spectral
energy distribution, 
which reflects the disk geometry and the structure of the dust
content. Disks have also
been observed more directly via optically thick dust lanes blocking the
scattered light from young stars, and as near-infrared images of the scattered
light of the disk itself \citep[e.g.][]{mccabe2002,weinberger2002}. 
Molecular line emission from species such as CO or HCO$^+$ is also used as
a tracer for disks. The use of such tracers is, however, subject to some
uncertainty. Heavy element molecules may freeze out on dust grains, which
likely settle to the midplane of the disk and/or get
bound in larger rocks or planetesimals. Thus, molecules such as CO can become
undetectable even if the disk still exists.

\subsection{Molecular Hydrogen in Disks}
Examining the H$_2$ content in disks has many advantages.
Hydrogen and helium are the last parts
of the gas to be bound up when planets form, and will therefore remain in the
disk after CO and dust have become undetectable. Observations of molecular
hydrogen directly trace the gas mass of the disk without making
assumptions about the dust-to-gas or CO-to-H$_2$ ratios.
Furthermore, molecular hydrogen
will remain in the surface layers of the disk when the dust settles to the
midplane and is more directly accessible to incoming light than the dust and
heavier elements. As a result, H$_2$ may prove to be a better tracer for
exploring the 
evolution and structure of circumstellar disks, since it may be observable for
a longer period of time.

Direct observations of H$_2$ in disks have been undertaken by several
groups. We focus here on the observations of the IR rovibrational lines,
although some studies have concentrated on pure rotational lines in the MIR
\citep[e.g.][]{lahuis2007} as
well as fluorescent H$_2$ in the UV \citep[e.g.][]{walter2003,herczeg2006}.   
Emission from the H$_2$ v=1-0 S(1) line at 2.1218$\mu$m has been detected in
the disks of several T Tauri stars, classical as well as weak-line
\citep{bary2003, bary2008,itoh2003,weintraub2005,ramsay2007,carmona2007}.  
These detections are made through longslit spectroscopic observations, and they do not
reveal much about the spatial distribution of the molecular hydrogen, beyond
indicating
that the emitting gas is located within 50 AU of the central
star. \cite{chen1998} presented images of H$_2$ v=1-0 S(1) emission from
photoevaporating 
disks in Orion and showed that the emission arises on the disk surface. In
this case, the disks were externally irradiated, and the H$_2$ emission was
found from a region $\sim 200$AU in size.

In this paper, we present spatially resolved images of H$_2$ emission from a
ring around T 
Tau N obtained with the integral field spectrograph SINFONI on the ESO-VLT. The
presence of H$_2$ emission in the T Tau system has been known for decades, but
this is the first time that the weak emission within 100 AU of T Tau N has
been resolved and analyzed.   

\subsection{T Tau}
T Tau is a triple star system with an age of $\sim 1$Myr
\citep{white2001}. The binary component T Tau S,
consisting of T Tau Sa and T Tau Sb (separation $\sim$0\farcs1), is currently $\sim$0\farcs7 south of T Tau
N. All three stars are actively accreting and believed to host disks
\citep{duchene2005}. T Tau S shows heavy extinction (A$_V=15$), which is
attributed to a circumbinary structure \citep{duchene2005}. Another
possibility is that T Tau S is obscured 
by the disk around T Tau N \citep{hogerheijde1997,beck2001}.  

T Tau N is a $\sim$2M$_{\odot}$ star \citep{white2001} and is believed to have a disk that is seen nearly face-on
\citep{akeson1998}. Based on 
photometric periodicity and assumed stellar radius, \cite{herbst1997} derive an
inclination of 
19$^{\circ}$. \cite{stapelfeldt1998} suggest an outflow and disk with the axis
at position angle 300$^{\circ}$ and with inclination of $\sim 45 ^{\circ}$ in
order to explain the morphology of scattered optical light. \cite{akeson2002}
find the inclination to be $20-40^{\circ}$ from SED fitting. 

This paper is organized as follows. In Sect.~\ref{obs}, we describe the
observations and data reduction. In Sect.~\ref{results}, we present the
spatial distribution of molecular hydrogen around T Tau N and the velocity
distribution of the gas. Section~\ref{geometry} discusses the geometry of
the star-disk-envelope system and Sect.~\ref{excitation} examines the H$_2$
excitation mechanism. In Sect.~\ref{ttaus}, we consider
the possible implications for T Tau S, and finally, we draw conclusions in
Sect.~\ref{conclusion}.

\section{Observations}
\label{obs}
T Tau was observed with the ESO-VLT as part of the SINFONI science
verification program on the 
nights of 2004 October 30th and November 2nd . SINFONI is a near-infrared
(NIR) integral field 
spectrograph working in combination with adaptive optics
\citep{eisenhauer2003}. Observations of the
region around the T Tau triple star system were obtained in the K-band using
the 3\farcs2 field of view optics (100 mas pixel 
scale) centered on the northern component. T Tau N (m$_V$=9.6) itself was used
as the guide star, producing
diffraction limited spatial resolution. 
The 2D image on the sky was sliced into 32 slitlets which were then dispersed
onto a 2k $\times$ 2k detector. 
The spectrograph provides a spectral resolution of 4000 in the K-band. The
observations were carried out using a five-point nodding pattern with
individual exposure times of 3 seconds and 20 co-adds and a total integration
time on source 
of 30 minutes. The nodding pattern was that of a box with one arcsecond width
  centered on T Tau N. The resulting mosaic has a field of view of
  approximately 4 \arcsec. 
Sky frames with the same exposure times were obtained 
within the nodding cycle. 

Data reduction and reconstruction of the 3D cubes were carried out using the
SINFONI pipeline (version 1.3.0) provided by ESO. The 2D raw frames were
corrected for sky 
background, flat field effects and optical distortions.
Bad pixels and cosmic rays were identified and the frames were calibrated in
wavelength. Then, the 3D cubes were constructed using calibration data of the
positions and distances of the slitlets on the detector. The cubes within the
nodding cycle 
were aligned spatially  and coadded plane by plane to create the final
  mosaic. Since the total exposure time is less at the outer regions of the
mosaic than in the centre we scaled the flux at all spatial points to an
exposure time of 3 seconds.  
The final 3D cube stores the spatial
information in the x- and y-directions and the spectral information along the
z-direction. To improve the signal-to-noise ratio, each spectral plane was
smoothed with a 3 by 3 boxcar in the spatial domain.

The B9 standard star Hip025657 was observed under the same conditions and
similar airmass as T
Tau and with the same instrumental setup, in order to correct for atmospheric
absorption. The spectrum was extracted after the
data had been reduced following the same recipe as for T Tau. The spectrum of
Hip025657 is featureless except for  Br$\gamma$ in absorption. We removed this
feature and replaced it by a linear fit to the surrounding
continuum. Subsequently, the spectrum was divided by a blackbody function of
T=11000 K and normalized. Dividing each spectrum of the science cube by the
corrected standard star spectrum removed 
telluric absorption features in the T Tau spectra very effectively.

Flux calibration was also performed using Hip025657 (m$_K$=7.443). The
conversion factor  
between counts s$^{-1}$ and ergs~s$^{-1}$ cm$^{-2}$ $\mu$m$^{-1}$ sr$^{-1}$~
was found by dividing the K-band flux 
of the star (4.1$\times 10^{-7}$ ergs s$^{-1}$ cm$^{-2}$ $\mu$m$^{-1}$~ $\times 
10^{-m_K/2.5}$ \citep{campins1985}) by the mean counts per second  of the
standard star spectrum within $1.94 - 2.45 \mu$m and 
dividing by the pixel area in steradians.

\section{Results}
\label{results}

   \begin{figure}
   \centering
   \includegraphics[width=0.5\textwidth]{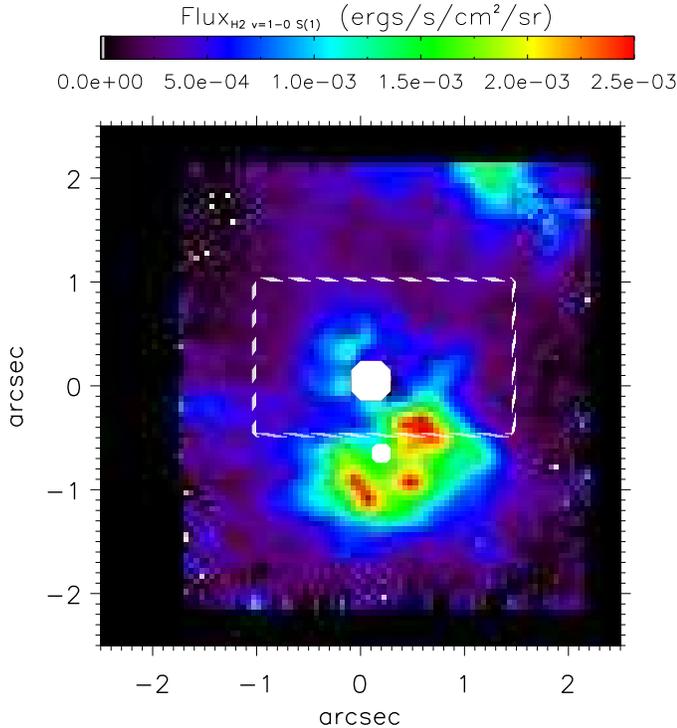}
      \caption{ H$_2$ v=1-0 S(1) emission in the T Tauri system. The color
        scheme indicate the flux level in ergs s$^{-1}$ cm$^{-2}$ sr$^{-1}$~. The positions of T Tau N
        and 
        T Tau S are marked with white circles. The white box outlines the
        close-up view of the ring-like structure around T Tau N shown in
        Fig.~\ref{fig:ring}.   
              }
         \label{fig:h2}
   \end{figure}

We show the spatial distribution of emission in the H$_2$ v=1-0 S(1)
rovibrational line at 2.12$\mu$m 
 in Fig.~\ref{fig:h2}. 
The image is dominated by strong H$_2$
emission south of T Tau N, close to T Tau S. This emission feature was also
detected by \cite{herbst2007} and \cite{beck2008}. 
The origin of the strong emission south of T Tau N is believed to be outflows
from one or more of the stellar components \citep{herbst2007} and will be the
subject of a subsequent paper (Gustafsson et al. in preparation).
In this paper, we focus on the weaker H$_2$ emission found very close to
  T Tau N. The weak feature is seen to
extend all around the star in a ring-like structure. 
The morphology of the H$_2$ emission in our data is fully
  consistent with the map recently published by \cite{beck2008}. They also
  detected the weak ring-like structure although they
  did not mention it. Their data
  were obtained October 2005, which indicates that the emission feature 
  is stable on at least a 1-year timescale.

\begin{figure*}
   \centering
   \includegraphics{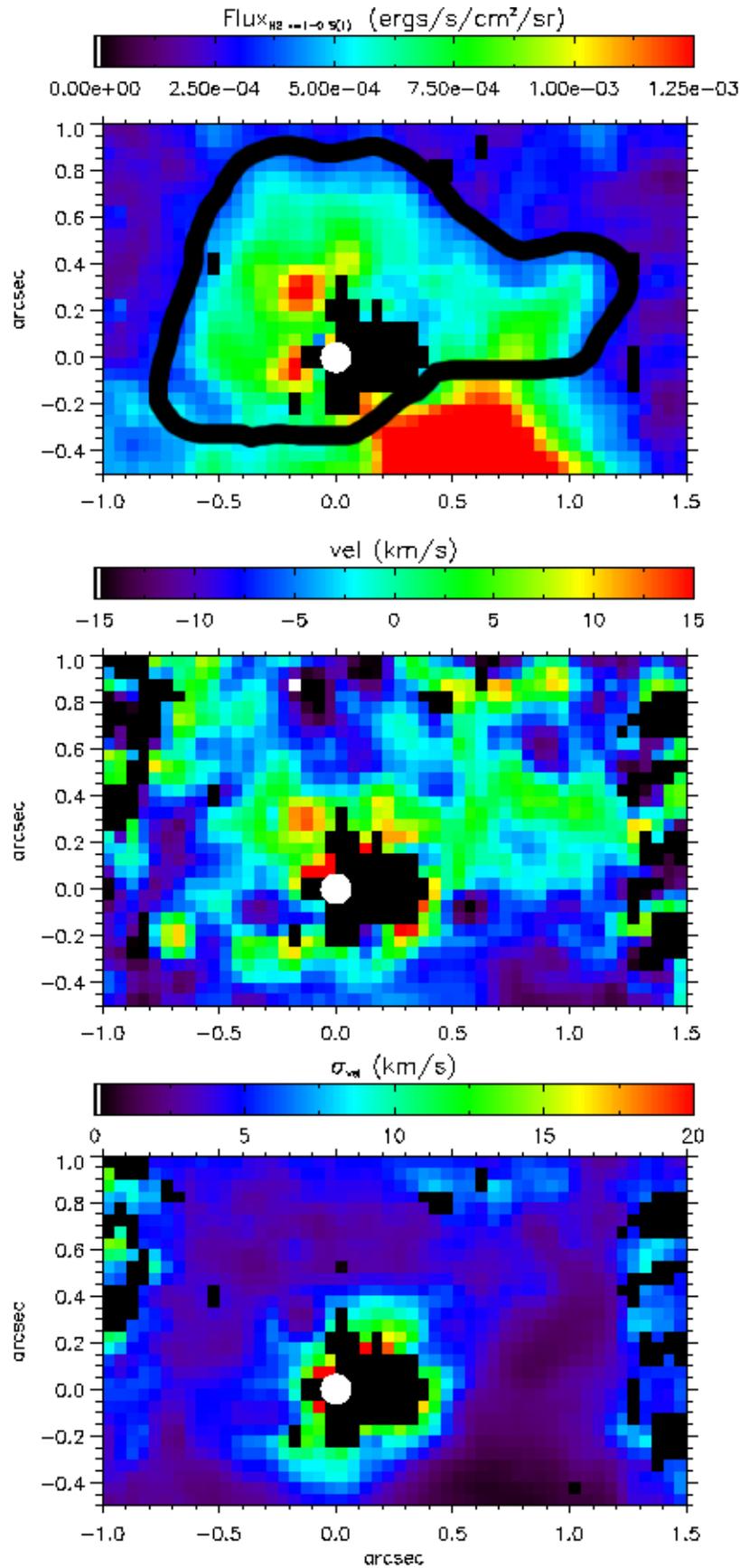}
      \caption{\small {Top: H$_2$ v=1-0 S(1) emission surrounding T Tau N. The
          color 
        scheme indicate the flux level in ergs s$^{-1}$ cm$^{-2}$ sr$^{-1}$~. Positions where no H$_2$
        emission is detected are shown in black and the position of T Tau N
        is marked by a white circle. The mask
        outlining the region used for further analysis is indicated by the
        black line. The strong emission region south-west of T Tau N is
        excluded because it is most likely related to outflows from T Tau S
        (see text). Centre: 
        The derived heliocentric radial velocity of the H$_2$ emission
        relative to the rest velocity of T
        Tau N. Bottom: Uncertainties in the radial velocities.}
              }
         \label{fig:ring}
   \end{figure*}

\begin{figure}
   \centering
   \includegraphics[width=0.5\textwidth]{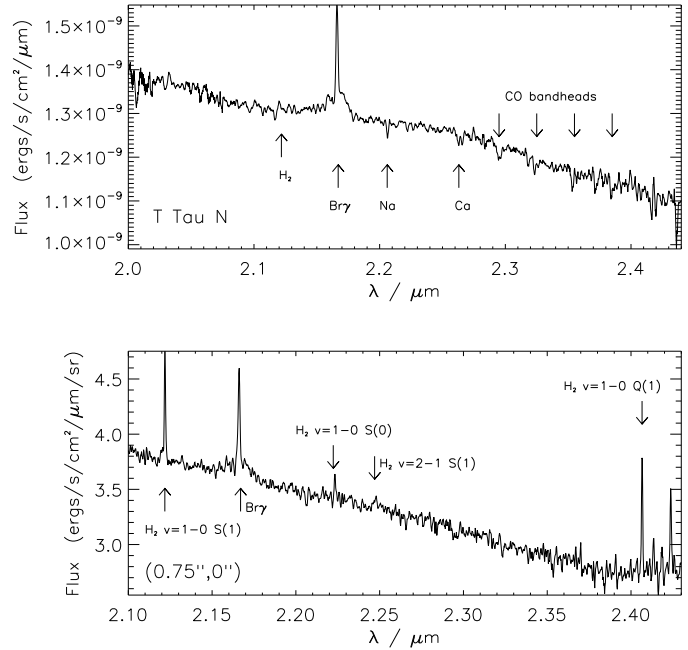}
      \caption{K-band spectrum of T Tau N and of a pixel within the mask in
        Fig.~\ref{fig:ring} located (0\farcs75,0)
        with respect to T Tau N. Br$\gamma$ 
        emission and photospheric absorption lines are seen in T Tau N. In the
      extended ring-like structure, only H$_2$ lines are detected. The
      Br$\gamma$ 
      line in the lower spectrum is contamination from the PSF of T Tau N and
      is not due to extended emission of Br$\gamma$.}
         \label{fig:spec}
   \end{figure}

A close-up of the immediate surroundings of T Tau N appears in
Fig.~\ref{fig:ring}.  
H$_2$ emission is found as close as 0\farcs1 arcsec ($\sim$ 15 AU, assuming a
distance of 140pc)
and is seen to extend out to $\sim$ 0\farcs7 (100 AU) from the star. We do not
detect H$_2$ emission above the noise closer to the star than
0\farcs1. Molecular hydrogen
does not appear in the spectrum of T Tau N (Fig.~\ref{fig:spec}) and is an
exclusively extended phenomenon. 

Emission from other H$_2$ lines than the 1-0 S(1) transition has also been
detected (Fig.~\ref{fig:spec}). The Q(1) line at 2.406$\mu$m is found to have
roughly the same 
spatial distribution as the S(1) line. Other Q-branch lines are present as
well, but the correction for atmospheric absorption is challenging
in this spectral and spatial region and may introduce errors. The S(0)
(2.2223$\mu$m) also 
appears in the spectrum. This line is, however, weak and 
the powerful continuum emission from T Tau N limits the line
detection at a level of 3$\sigma$ above the noise level to pixels located in
the outer ring structure at r$>70$ AU. At larger radii, the  S(0) emission
shows a similar spatial distribution to the S(1) line. The v=2-1 S(1) line
(2.2447$\mu$m) is tentatively 
detected at a 2$\sigma$ level at a few locations in the outer region of the
ring structure at r$>80$ AU.  

The total flux of the H$_2$ v=1-0 S(1) in the structure is found by summing
all light within a mask, the extent of which is shown in Fig.~\ref{fig:ring}.
We have chosen a rather conservative mask in order to avoid confusion with
emission features that may have a different origin. Thus,
the mask excludes regions where the emission is weaker than $4 \times 10^{-4}$
ergs s$^{-1}$ cm$^{-2}$ sr$^{-1}$~ as well as the strong emission region south-west of T Tau N which
is most likely caused by an outflow from one of the T Tau S stars
\citep{herbst2007}. 
The total flux within the mask is $1.8 \pm 0.3 \times 10^{-14}$ ergs s$^{-1}$ cm$^{-2}$~. The
uncertainty 
is the formal uncertainty on the total flux calculated using the flux
uncertainty 
in the pixels included in the sum.
This uncertainty does not take into account
that the estimated flux depends on the chosen mask and that the exact shape
and extent of the ring-like structure is difficult to quantify because of
other H$_2$ features nearby. We estimate that the uncertainty due to the
mask may amount to 30\%. 
The flux of the v=1-0 Q(1) line and the upper 
limits to the flux of other H$_2$ lines appear in Table~\ref{tab:flux}.

\begin{table}
\caption{Total flux of H$_2$ lines within mask.$^a$: 3$\sigma$ upper limit.}
\begin{tabular}{cccc}
\hline
line& $\lambda$ &flux &ratio $F_{v=1-0  S(1)}/F_{line}$\\
 & ($\mu$m)&($10^{-14}$ergs s$^{-1}$ cm$^{-2}$~)&\\
\hline
v=1-0 S(1) & 2.1218&1.8 $\pm 0.3$ & 1.0\\
v=1-0 S(0) & 2.2223&$<$0.5$^a$ & $>$3.6\\
v=2-1 S(1) & 2.2477&$<$0.6$^a$ & $>$3.0\\
v=1-0 Q(1) & 2.4066&1.9 $\pm 0.1$ & 0.9\\
\hline
\end{tabular}
\label{tab:flux}
\end{table}

The total flux in the v=1-0 S(1) line is similar to the amount of
H$_2$ emission at the same radial velocity as the star 
detected in the circumstellar environment of other T Tauri stars, where it is
believed to originate from disks within 100 AU \citep{bary2003,bary2008,weintraub2005,
  ramsay2007}. The H$_2$ line
flux previously measured in disks ranges from $7\times 10^{-16} - 1.5\times
10^{-14}$ergs s$^{-1}$ cm$^{-2}$~.

   \begin{figure}
   \centering
   \includegraphics[width=0.5\textwidth]{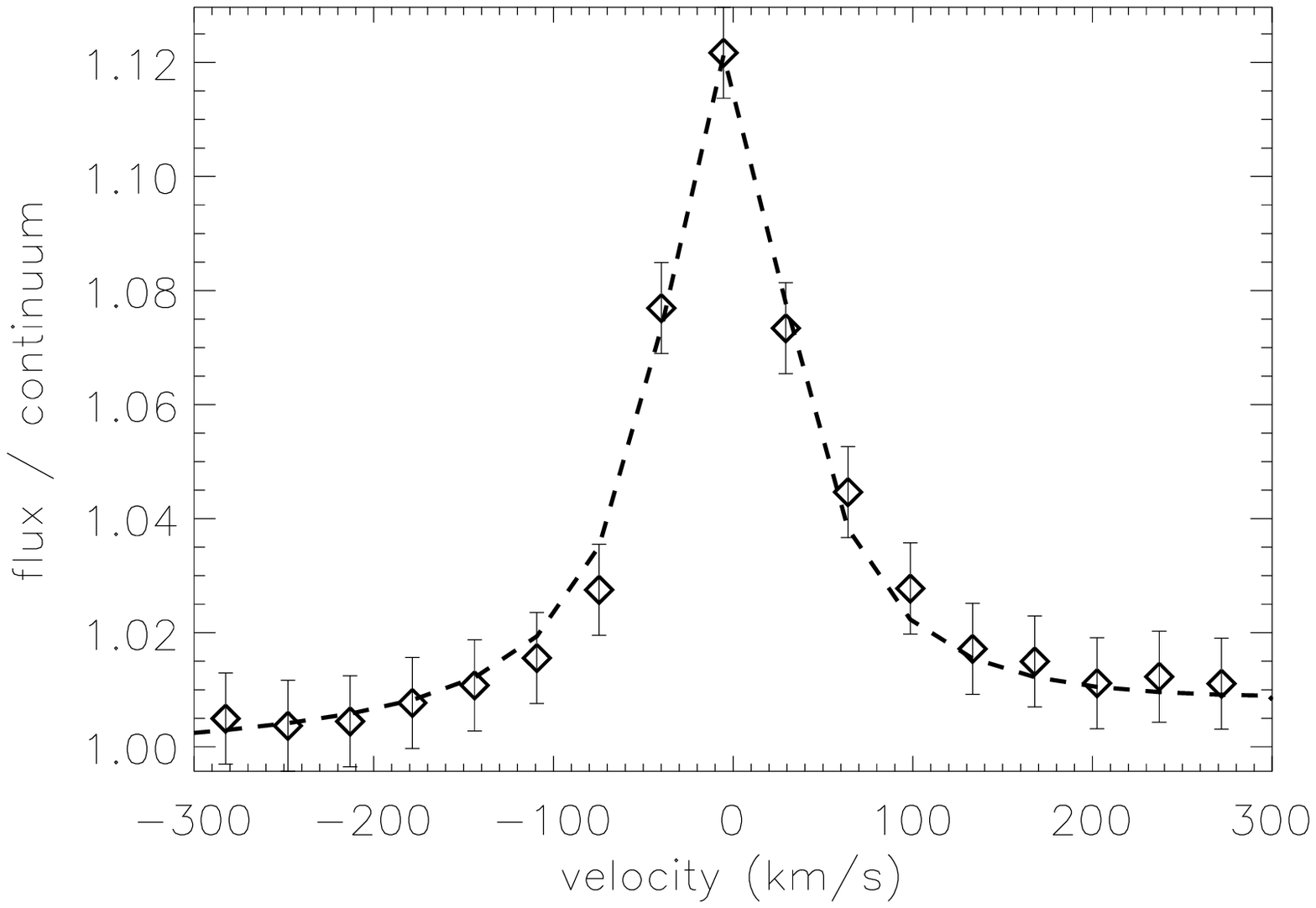}
      \caption{ H$_2$ v=1-0 S(1) line profile integrated over the ring structure
(diamonds with 1$\sigma$ errorbars). Fitted Lorentzian (dashed
line). }
         \label{fig:velprof}
   \end{figure}

In our data, we have full access to the spectral distribution of the emitting
gas. Although SINFONI only has a spectral resolution of $\sim 75$
kms$^{-1}$~ in the K-band, it is possible to determine the peak position of the lines
with much higher accuracy through line fitting. We have derived the radial
velocity 
corresponding to every H$_2$ emitting position (see Fig~\ref{fig:ring}) by
fitting a Gaussian profile to the unresolved line profiles on a pixel by pixel
basis. The
velocities in Fig.~\ref{fig:ring} have been corrected for the Earth's motion
toward T Tau at the 
time of observation and are quoted with respect to the heliocentric
velocity of T Tau 
N of $ 19.1 \pm 1.2 $kms$^{-1}$~ \citep{hartmann1986}. The
velocity map in Fig.~\ref{fig:ring} shows small velocity variations between
-10kms$^{-1}$~ and +10kms$^{-1}$~ with 
respect to the intrinsic velocity of T Tau N within the ring structure. 
These velocities are consistent with the data from \cite{beck2008} who also measure velocities close to the systemic velocity
  at this location (see their figure 11).
There is a tendency for a 
radial gradient with the velocities being positive (5-10kms$^{-1}$~) close to
the star and 
negative (-5kms$^{-1}$~) further out. However, the uncertainty in the derived
velocities are 
larger close to the star than further out due to the increased continuum
emission in
the inner region (Fig.~\ref{fig:ring}). This makes any conclusion on
the radial variation of velocities rather uncertain.   
There is no evidence of Keplerian rotation of the disk. However, if the
inclination of the disk is the same as the star itself, $\sim 20^{\circ}$, the
radial velocity component of 
Keplerian rotation around a
2M$_{\odot}$ star is only $\sim 5$kms$^{-1}$~ at 10AU and $\sim 2$kms$^{-1}$~ at 100AU. Such
small velocity differences within the disk would be difficult to detect with
the present data.

In order to improve the signal-to-noise ratio, we constructed a global H$_2$
profile of the ring-like structure by adding all spectral profiles of H$_2$
v=1-0 S(1) emitting positions within the mask in Fig.~\ref{fig:ring}. This
also allows a direct comparison with previous 
spatially unresolved measurements of H$_2$ in the circumstellar environment of
T Tauri stars. The global line profile appears in Fig.~\ref{fig:velprof}
together with a Lorentzian fit. The Lorentzian fitting function provides the
best match to the instrumental profile of SINFONI which
dominates the unresolved H$_2$ profile.
The profile is seen to peak
close to the rest velocity of T Tau N. From the Lorentzian fit we find the peak
velocity to be $-2.5 \pm 2.1$kms$^{-1}$~ ($1\sigma$ uncertainty). Considering the
uncertainty  in the rest velocity of T Tau N of 1.2kms$^{-1}$~ \citep{hartmann1986},
the velocity of the H$_2$ emission 
is consistent with the rest velocity of the star within the errors. The same
was found to be true of the H$_2$ emission from disks around other stars
\citep{bary2003, bary2008,ramsay2007,carmona2007}.

\section{Disk or outflow?}
\label{geometry}

We now examine the origin of the H$_2$ emission around T Tau N.
The ring-like shape of the emission can be created by several scenarios. We 
consider the following possibilities: 1) an outflow, 2) shocks
created by 
a wide-angled wind hitting the disk or cavity walls, 3) a photo-evaporating
wind from a disk viewed almost face-on \citep{hollenbach1994,font2004}, 4) a
photodissociation region, or 5) UV/X-ray 
heating of a disk \citep{nomura2007}. In this section, we consider the
geometry of the system and in Sect.~\ref{excitation}, we examine what
excitation mechanisms can reproduce the H$_2$ line flux and ratios.

\begin{figure}
   \centering
   \includegraphics[width=0.5\textwidth]{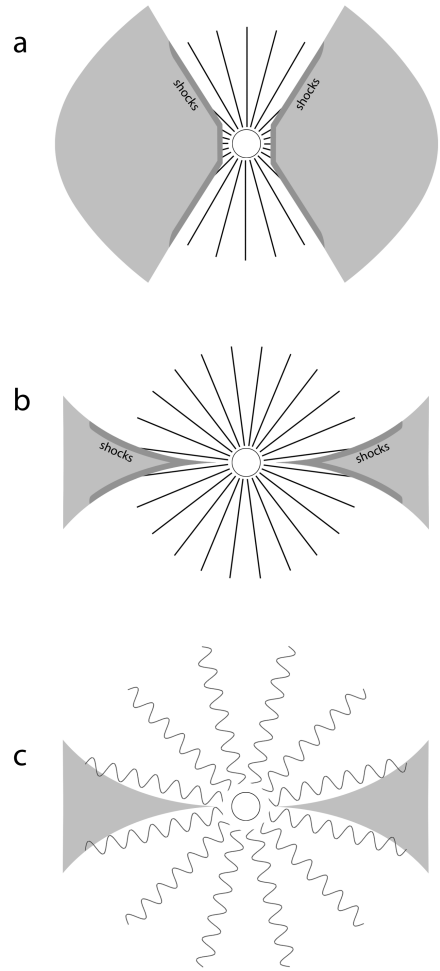}
      \caption{Excitation scenarios in the T Tau N-disk-envelope system. {\bf
          a):} a wide-angle outflow hits the walls of a cleared out 
      cavity in the surrounding envelope, {\bf
          b):} a wide-angle outflow impinges on a flared disk and creates
        shocks, {\bf
          c):} UV + X-ray irradiation from T Tau N heats a flared disk.}
         \label{fig:models}
   \end{figure}

Outflows from T Tauri stars are known to be complex, often showing two velocity
components with different spatial characteristics. Observations of both
forbidden 
emission lines \citep{bacciotti2000} and molecular hydrogen \citep{takami2006}
have shown that outflows typically consist of a collimated high velocity
(60-200kms$^{-1}$~) jet and a less collimated low velocity
(0-30kms$^{-1}$~) component, i.e. a wide-angled wind.
 At first glance, the small line-of-sight velocities measured in the ring argue
against a collimated jet. 
Although a jet viewed almost pole-on could create a circular
emission feature, the velocities associated with such a jet 
are much higher than observed here. It
is, however, possible that we see shocks 
from a wide-angle wind interacting with either the outer walls
of a bi-conical cavity cleared out by the outflow or a flared disk
(Fig.~\ref{fig:models}a or~\ref{fig:models}b ). The 
existence of a cavity in an envelope around T Tau N was 
suggested by \cite{momose1996} 
and the reflected light images of \cite{stapelfeldt1998}. 
In this picture, a wide-angle wind creates oblique shocks when interacting with
the molecular environment. Oblique shocks result in low shock velocities,
since it is only the normal component of the velocity with respect to the gas
that is thermalized. Furthermore, the measured velocities will be even lower,
since only the line-of-sight velocity is detected here.

Another possibility is that the emission is linked to irradiation of a nearly
face-on disk around T Tau N (Fig.\ref{fig:models}c).  
We proceed by investigating the circumstellar environment of T Tau N through
spectral energy distribution (SED) modeling to find evidence of the size of
the disk and envelope. We use the precomputed grid of
radiative transfer models of \cite{robitaille2007}. Their models include
contributions from a circumstellar accretion 
disk, an infalling envelope and an outflow cavity. The models span a wide
range of 
parameter space. The SED is calculated for each set of model parameters for
comparison  with observational data. Using fluxes of T Tau N from the
literature, we fit the 
optical and infrared data with the models in order to 
find the range of physical parameters that fit the SED best.

\begin{figure*}
   \centering
   \includegraphics[width=0.7\textwidth]{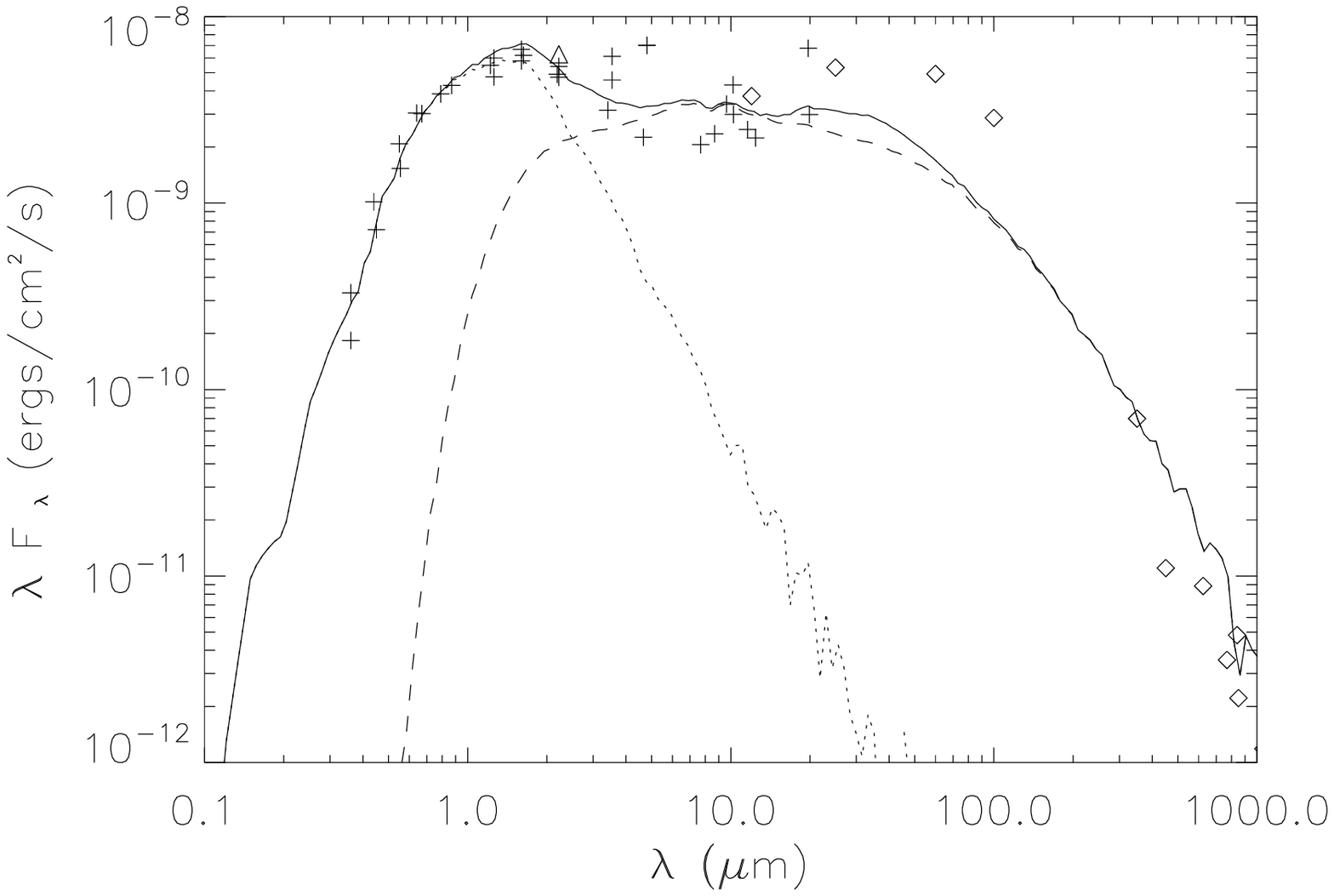}
      \caption{ Spectral energy distribution of T Tau N and model curves. The
        data points include those of this work and
        \cite{herbst1997,herbst2007,kenyon1995, 
          hogerheijde1997,weaver1992,andrews2005,beckwith1991}. 
        Plus-signs indicate resolved data of T Tau N, diamonds indicate
        where T Tau S and T Tau N are unresolved and the flux of T
        Tau S dominates. The K-band flux from this
        work is marked with a triangle. The model
        is found from fitting to data at $\lambda < 21 \mu$m where T Tau N and
        T Tau S are resolved. Solid
        line: total flux, dotted line: stellar flux, dashed line: disk flux. }
         \label{fig:SED}
   \end{figure*}

The SED appears in Fig.~\ref{fig:SED}. The observations include those of 
\cite{herbst1997,herbst2007,kenyon1995,hogerheijde1997,weaver1992,andrews2005,beckwith1991}. 
T Tau N shows brightness varitations of 0.2 mag in the K and L bands
\citep{beck2004}, which can explain the spread in flux values that the 
various authors find. We have included measurements from different epochs at
optical and infrared wavelengths, in order to average out the effect of this
variability.   
In the model fit, we have only used data points at $\lambda < 21\mu$m, for
which 
the T Tau N component is resolved. At longer wavelengths, the north and south
components of the T Tau triple system are unresolved and T Tau S
dominates \citep{ghez1991}. We assume that A$_V = 1.5$ \citep{white2001}. 

The Robitaille grid resulted in 
a wide range of parameters that produced a reasonable fit to the SED. In order
to further constrain the parameter space, we used the spectral information on T
Tau N,  a $\sim1$Myr old K0 star with M$\sim2$M$_{\odot}$,
T$_{\rm eff}=5200$K \citep{white2001}. Thus, we chose to consider only those
models in which the stellar parameters fall within the following ranges, 
$1.5<$M/M$_{\odot}<$2.5, 4000K$<$T$_{\rm eff}<$6000K, 0.1Myr$<$age$<$2Myr. The
best fit model which satisfies these criteria appears in Fig.~\ref{fig:SED}.
For the 50 models that best fit the data, only four models satisfy
the stellar restrictions of T Tau N. These four models have the same values
for all parameters except the
inclination angle of the disk. The model parameters are listed in
Table~\ref{tab:SEDmod}. The inclination is poorly determined, but is found to
be less than 65$^{\circ}$.  

\begin{table}
\centering
\caption{SED model parameters}
\begin{tabular}{cc}
\hline
envelope accretion rate& $10^{-7}$M$_{\odot}$/yr\\
cavity opening angle& $47^{\circ}$\\
Disk mass & 0.15 M$_{\odot}$\\
Disk outer radius & 85 AU\\
Disk inner radius & 0.3 AU\\
flaring parameter, $\beta$& 1.14\\
Disk accretion rate & $10^{-6}$M$_{\odot}$/yr \\
Disk inclination & $<65^{\circ}$\\
\hline
\end{tabular}
\label{tab:SEDmod}
\end{table}

SEDs only contain a limited amount of information about the distribution of
circumstellar material, and it is easy
to overinterpret the results of model fitting. Nevertheless, our
results show that
the T Tau N system is consistent with an accreting star surrounded by a
disk and torus-like envelope with an opening angle of $\sim 45^{\circ}$. The
disk is likely to have an outer radius of 85 AU consistent with the detection
of H$_2$ at a radius of $\sim 100$AU. It also shows a large degree of flaring.  
The inclination of the disk cannot be well constrained by the SED-modeling
but is consistent with a low inclination of $\sim19^{\circ}$, as found by
\cite{herbst1997}. 
The model is also consistent with the suggestion of
\cite{stapelfeldt1998} that an outflow at PA=300$^{\circ}$ and inclination of
45$^{\circ}$ has blown out a cavity. Note that the disk accretion rate is
higher than the observed accretion rate of $(1.4-5.9) \times
10^{-8}$M$_{\odot}$ yr$^{-1}$ 
based on the Br$\gamma$ line strength \citep{beck2004}.
This is most likely due to inconsistencies in the radiative transfer models
used in the Robitaille models, which systematically overestimate the accretion rate. This effect was already
noted in \cite{robitaille2007}. Thus, the accretion rate inferred from the
SED modeling should be considered with some caution.

Assuming optically thin emission, the mass of H$_2$ v=1-0 S(1) emitting gas is \citep{bary2003}
\begin{equation}
M(H_2)_{v=1-0 S(1)}= 1.76 \times 10^{-20} \frac{4\pi F_{line}
  D^2}{E_{v,J}A_{line}}=5\times 10^{-10} M_{\odot},
\label{eq:h2mass}
\end{equation}
where  $F_{line}$ is
the v=1-0 S(1) flux, D the distance in pc, $E_{v,J}$ the 
upper level energy and $A_{line}$ the Einstein A coefficient of the
transition.   
Following \cite{bary2003}, we estimate the total disk mass by multiplying the
v=1-0 S(1) H$_2$ mass by a scaling factor of $10^7-10^9$. 
This scaling factor was found by comparing the H$_2$ mass to masses derived from
CO and (sub)millimeter continuum in disks around four T Tauri
stars (GG Tau, LkCa 15, DoAr 21, TW Hya). 
This gives a disk mass of $0.005-0.5$M$_{\odot}$. Note that the most
likely disk mass of 0.15M$_{\odot}$ estimated from
the SED modeling falls within this range. 
Analysis of HCO$^+$ and millimeter continuum data indicate that the disk mass
around T 
Tau N is $ 10^{-3} - 4\times 10^{-2}$ M$_{\odot}$
\citep{hogerheijde1997,akeson1998}, suggesting that the actual disk mass is
at the lower end of the simple mass estimate based on the H$_2$ mass.

These arguments suggest that T Tau N has a flared disk with radius 85-100
AU and mass $0.005-0.5$M$_{\odot}$. We propose that the
observed H$_2$ emission comes from the gaseous disk atmosphere or
alternatively from the 
torus-like envelope that is also present around T Tau N according to the SED.
The reason that T Tau N is not at the centre of the H$_2$ emission (Fig.~\ref{fig:ring}) could be
due to the fact that we do not see the disk-envelope system exactly
face-on. An inclination of 20-40$^{\circ}$ would cause emission from a
circular structure to appear elliptical. Assuming that we only see emission
from the surface facing toward us projection effects would furthermore make
the nearest side of a disk appear narrower than the farther side. This
would shift the projected position of the central star away from the centre of
emission. A detailed modeling of the spatial
distribution of the emission is, however, outside the scope of this paper.

If the value of 85-100 AU reflects the true size of the disk,
why has the disk not yet been detected in scattered light?
The simplest explanation is that the disk is small (r $<$ 0\farcs7) and the
flux contrast between the star and disk is large, making detection
difficult. Infrared
observations of scattered light from nearly face-on disks (circumstellar and
circumbinary) around other T Tauri stars have shown that the total flux from
a disk is only 1-2 percent of the stellar flux \citep{mccabe2002,
  weinberger2002}. 
\cite{stapelfeldt1998} and \cite{mayama2006} observed reflected light in T Tau
at optical and infrared wavelengths, respectively, but did not find evidence
of the disk around T Tau N. However, \cite{mayama2006} used a coronographic
mask 
with a diameter of $\sim$0\farcs5 -0\farcs6, which is only marginally smaller
than the size of the disk found in this paper. The disk would therefore in any
case be difficult to see in their data. The detection of the disk from the
data of \cite{stapelfeldt1998} could be compromised by artefacts arising from
the PSF subtraction.

\section{Excitation mechanism}  
\label{excitation}

After studying the location of the circumstellar material and the H$_2$
emitting gas, we now turn to the H$_2$ excitation mechanism. 
The near-infrared rovibrational lines of H$_2$ can be excited through different
processes. The two main mechanisms are (i) shocks (Fig.\ref{fig:models}a and
\ref{fig:models}b) and (ii)
UV + X-ray radiation (Fig.\ref{fig:models}c). In the former case, the H$_2$
molecules are thermally 
excited by the passing shock wave and the H$_2$ spectrum is characterized by a
single excitation temperature. In the latter case, the UV and X-ray radiation
may contribute both to the heating of the gas and to electronic excitation of
H$_2$. The near-infrared H$_2$ emission may therefore
contain a contribution from both thermal excitation as well as non-thermal
radiative decay from excited electronic states (fluorescence). 
These two excitation mechanism can be distinguished through both the line
strengths and the line ratios of the
observed H$_2$. Traditionally, the ratios $I_{v=1-0 S(1)}/I_{v=1-0 S(0)}$
and $I_{v=1-0 S(1)}/I_{v=2-1 S(1)}$ are used. These ratios also provide a
diagnostic for whether the thermal or non-thermal contribution dominates,
since higher vibrational bands, $v \ge
2$, are more densely populated by fluorescence than by thermal excitation. 
In the following, we will concentrate on the two excitation processes (i)
shocks and (ii) UV + X-ray irradiation

\subsection{Shocks}
In the first process, shocks, a super-Alfvenic shock wave rapidly heats the gas
to temperatures of $\sim 1000 - 3000$K. The temperature reached depends on the
velocity of the shock, as well as the pre-shock conditions in the ambient
medium, such as density, magnetic field, chemistry etc. The $I_{v=1-0
  S(1)}/I_{v=2-1 S(1)}$ ratio depends on the temperature, since the populations
are assumed to be in local thermodynamic equilibrium (LTE). The higher the
temperature, the larger the relative population in higher vibrational bands. 
We estimate the range of the  $I_{v=1-0 S(1)}/I_{v=2-1 S(1)}$ ratio that
shocks are able to produce considering the temperature range 1000
-3000K reached during the passing of the shock. For an optically thin
transition, the line intensity is  
\begin{equation}
I_{line}= \frac{A_{line}hc}{4 \pi \lambda} N_{v,J},
\end{equation}
where $N_{v,J}$ is the population in the upper level denoted by the v and J
quantum numbers, $\lambda$ the wavelength,
and $A_{line}$ is the Einstein coefficient of the transition. We ignore
extinction effects, since the differential extinction between the H$_2$ lines
is small.
In LTE, the level population is given by
\begin{eqnarray}
N_{v,J}=\frac{N_{tot}}{Z(T)}g_{v,J}\exp(-E_{v,J}/T),
\end{eqnarray}
where $N_{tot}$ is the total column density, $Z(T)$ is the partition function,
$g_{v,J}$ is the degeneracy of the level, $E_{v,J}$ is the level energy
expressed in K, and T is the excitation temperature.
The degeneracy depends only on the rotational state, J. Thus, $g_{1,3}
= g_{2,3}$.  
We find:
\begin{eqnarray}
\frac{I_{v=1-0 S(1)}}{I_{v=2-1 S(1)}}&=&\frac{A_{v=1-0 S(1)}}{A_{v=2-1
    S(1)}}\frac{2.24\mu m}{2.12\mu m} exp \left(
  {\frac{E_{2,3}-E_{1,3}}{T}} \right) \nonumber \\
&=& 0.736 exp\left (\frac{5594 K}{T} \right),
\end{eqnarray}
which gives a value of 198 at 1000 K and 5 at 3000K.
Similarly, we find $I_{v=1-0 S(1)}/I_{v=1-0 S(0)} \simeq 3-5$ for T$\sim 1000
-3000$K.

From these estimates, it is clear that a range of shock temperatures can
reproduce the line 
ratios observed around T Tau N (Table~\ref{tab:flux}). However, since the
observations give us only lower limits to the line ratios, we are not able to
constrain the shock temperature and thus the underlying physical parameters
very well. We therefore turn to the total line flux. 
Shock models show that the H$_2$ v=1-0 S(1) surface brightness in 
Fig.~\ref{fig:ring} can be produced in different
density regions of the circumstellar material. The required density is tightly
correlated with the impact velocity of the 
shock front. This means that the H$_2$ emission can arise in environments with
pre-shock density of n$_H \sim 1\times10^{4}$cm$^{-3}$ if the shock velocity
is $\sim 40-50$\kms while a shock front propagating at only $\sim 10$\kms is
sufficient to excite the H$_2$ if the pre-shock density is n$_H \sim
1\times10^{7}$cm$^{-3}$ (L. E. Kristensen, private communication).

From models of T Tauri stars with disk and envelope masses resembling those of
T Tau N, it is evident that the high 
density regime of n$_H \sim 
1\times10^{7}$cm$^{-3}$ is associated with the upper layers of the disk while
densities of   
n$_H \sim 1\times10^{4}$cm$^{-3}$ are found in the envelope and in a thin
transition zone between the disk atmosphere and the outflow cavity, i.e. the
cavity walls \citep{whitney2003, crapsi2008}. Independent of where the
excitation 
takes place, the outflow or wind must impact the gas at a small angle, unless
the disk is highly flared. Thus, the flow velocity
must be much higher than the shock velocity, since it is only the velocity
component perpendicular to the shock surface that contributes to the
shock. Assuming an 
impact angle of 10$^{\circ}$-20$^{\circ}$, a flow velocity of 120-230\kms is
required to create a shock of 40\kms. Such velocities are found in the
collimated jets from T Tauri stars. However, the SED indicates that the cavity
opening angle is large ($\sim 47^{\circ}$, Table~\ref{tab:SEDmod}) and thus it
is difficult to see how 
a collimated jet should be able to hit the cavity walls. On the other hand, a
shock velocity of 10\kms can be attained with a flow velocity of 30-60\kms and
an impact angle of 10$^{\circ}$-20$^{\circ}$. Such velocities may be found in
a wide-angled low velocity wind characteristic of T Tauri stars
\citep{bacciotti2000, takami2006}. 
 
In summary, if shocks are the primary cause of 
the H$_2$ excitation, it seems more likely that they arise from the
interaction of a wide-angled wind with the upper layers of a flared disk than
with cavity walls carved in a circumstellar envelope. We cannot
exclude that the H$_2$ excitation arises from the cavity walls, but this
scenario seems to require that the high velocity jet in T Tau N be less
collimated than in other sources.

\subsection{Irradiation}
We now turn to excitation process (ii), irradiation by UV-photons and X-rays
(Fig.\ref{fig:models}c).  
In their models of fluorescent excitation of H$_2$, \cite{black1987} find 
$I_{v=1-0 S(1)}/I_{v=1-0 S(0)} < 2.7 $
and $I_{v=1-0 S(1)}/I_{v=2-1 S(1)} < 2.0$. This is incompatible with our
data (Table~\ref{tab:flux}) and shows that fluorescence alone cannot explain
the observations. A thermal component is necessary in order to reproduce the
line ratios. 
\cite{lepetit2006} have shown that the
ratio $I_{v=1-0 S(1)}/I_{v=2-1 S(1)}$ can be much larger than 2 in a
photon-dominated region (PDR) if n$_H > 10^5$cm$^{-3}$ and the incident
far-ultraviolet radiation field is stronger than $\sim 10^4$ times the average
in the interstellar medium. In such high density regions, collisions of H$_2$
in vibrationally excited states resulting from fluorescence with atomic H tend
to thermalize the rovibrational states. 
These conditions may very well apply to a dense
circumstellar disk irradiated by the central star.
The brightness of the v=1-0 S(1) line in the model is found to be larger than
$\sim 5\times 10^{-4}$ergs s$^{-1}$ cm$^{-2}$ sr$^{-1}$~ 
in a PDR viewed face on \citep{lepetit2006}, which is consistent with our
data. The H$_2$ brightness around T Tau N is larger than this
value out to distances of $\sim 80$AU 
(Fig.~\ref{fig:ring}). The H$_2$ emission  
may thus be caused by UV-irradiation from T Tau N creating a PDR at the dense
surface 
of the disk. Such a PDR would create fluorescent H$_2$ lines in the UV and
infrared H$_2$ lines composed of contributions from both fluorescence and
thermal excitation.   
\cite{walter2003} found extended fluorescent H$_2$ emission in
the UV around T Tau N. This seems to support our
conclusion. \cite{herczeg2006}, however, did not find fluorescent H$_2$ emission
in the UV extending more than 0\farcs1 from T Tau N.
\cite{saucedo2003} reported fluorescent H$_2$ emission North-East and
South-West of T Tau N at a distance of $\leq 50$ AU, which is consistent with
pumping by stellar Ly$\alpha$ emission. They do, however, find that the
fluorescent line needs to be heated before being pumped and suggest that an
outflow must be the heating mechanism.

\cite{nomura2007} constructed models of H$_2$ emission from a disk irradiated by
both X-rays and UV photons from a central T
Tauri star. For a model with a 0.5M$_{\odot}$, T$_{eff}=4000$K central star
with X-ray luminosity of L$_X 
\sim 10^{30}$ergs s$^{-1}$ and UV excess like TW Hydrae, they find v=1-0 S(1)
line 
fluxes of $0.1-20\times 10^{-15}$ ergs s$^{-1}$ cm$^{-2}$~, depending on the size of the dust
grains. When the grain size increases the high-temperature region in the disk
shrinks and the line flux decreases.
The $I_{v=1-0  S(1)}/I_{v=2-1 S(1)}$ ratio is always larger than 4 and 
can be as high as 50 if UV radiation dominates. 
In units of $10^{-14}$ergs s$^{-1}$ cm$^{-2}$~, the estimated flux in the v=1-0 S(1) line for
a model using small dust grains is
found to be $\sim 0.1$ 
when only X-ray irradiation is considered, $\sim 1.3$ with only
UV-irradiation and $\sim 2.1$ when both X-rays and UV-irradiation are included
\citep{nomura2007}. 
The estimated total flux from the disk around T Tau N is 1.8 on this scale,
suggesting that X-rays alone are not sufficient to produce
the observed H$_2$ emission. UV-photons seem to dominate the excitation but a
combination of X-rays and UV-photons may be necessary in order to explain the
H$_2$ flux as emission from a disk around T Tau N. 
Note that T Tau has an X-ray luminosity of L$_X= 3.5 \times 10^{31}$ergs
s$^{-1}$ 
\citep{gudel2007} and UV-luminosity of L$_{UV}= 3 \times 10^{33}$ergs s$^{-1}$,
corresponding to a large UV-excess of 0.7L$_{\odot}$ \citep{calvet2004}.

It seems likely, then, that the UV-irradiation
from T Tau N itself, with a possible X-ray contribution, is strong enough to produce the total H$_2$ flux we see
around T Tau N. However, 
in their models, \cite{nomura2007} find that most H$_2$ emission in the v=1-0
S(1) line is emitted at a radius of 20AU and that the emissivity decreases
with increasing distance from the star. The presence of H$_2$ emission
extending to $\sim 100$AU around T Tau may be difficult to explain with models
of irradiation. A possible way to solve this is if
the disk is strongly flared. A large degree of flaring will allow more
UV-photons to reach the outer parts of the disk, thus increasing the excitation
in the outer regions.

\subsubsection{Photoevaporation}

If the disk is irradiated by UV photons and X-rays, a disk-wind may
be powered 
by photoevaporation of the disk
\citep{hollenbach1994,johnstone1998,storzer1999,font2004}. The high-energy
radiation (EUV 
+ X-rays) ionizes hydrogen at the disk surface and heats the gas to
$10^4$K. EUV photons cannot penetrate the ionization front and X-rays 
heat only the inner disk and the surface layer \citep{nomura2007}.
FUV photons, which are not absorbed by atomic hydrogen but mainly by dust,
penetrate much deeper and reach the disk surface. Here, they dissociate
molecular hydrogen and heat the neutral gas to about 400-4000 K. If the
thermal velocity of the neutral gas exceeds the escape velocity of the disk
surface, the gas flows outward. The flow is initially cylindrical but is
reoriented into a spherical flow by pressure gradients
\citep{font2004}. Deeper in this photon-dominated region, 
there is an H/H$_2$ transition layer below which the gas is molecular. The
H$_2$ v=1-0 S(1) line is emitted from the H/H$_2$ transition layer, which is
close to the disk surface \citep{storzer1999}. See Fig.~13 in
\cite{dullemond2007} for an illustration of the structure of a
photoevaporative disk.
\cite{storzer1999} pointed out
that the ionization structure in a PDR with photoevaporation may differ from a
classical 
PDR because the material is not at rest. They also found that the H$_2$ v=1-0
S(1) and v=2-1 S(1) lines are mainly collisionally excited, and that
fluorescence contributes only a small amount to the line intensities. The 
$I_{v=1-0  S(1)}/I_{v=2-1 S(1)}$  ratio is typically 5-10. 

If the molecular hydrogen in the H/H$_2$ transition layer gets hot enough to
overcome the escape velocity it may contribute to the gas flow away from
the disk. This could explain the small blueshift of the line
profile in Fig.~\ref{fig:velprof}.

\subsubsection{Mass loss}
It is interesting to consider the scale of the mass loss due to
photoevaporation. 
A characteristic radius for thermal evaporation is \citep{dullemond2007}:
\begin{equation}
r_g \sim 100 {\rm AU} \left( \frac{T}{1000K}\right)^{-1} \left(\frac{M_{\star}}{M_{\odot}}\right),
\end{equation}
where the sound speed equals the escape speed. However, simulations have shown
that evaporation is initiated at smaller radii than this analytical result
suggests. That is, photoevaporation happens outside $r_{cr}> 0.15r_g$
\citep{dullemond2007}. Using M=2M$_{\odot}$ and T=$10^4$K as a typical
temperature of the ionized gas, we find $r_{cr} \sim 3$AU or about 20 mas. It
thus seems very likely that the H$_2$ emission caused by irradiation is
accompanied by a 
photoevaporative wind although it is not directly detected in these
observations.

\cite{hollenbach1994} derived an estimate for the hydrogen mass loss rate due
to photoevaporation, 
\begin{equation}
\dot{M}_{wind} \simeq 4\times 10^{-10} \left(\frac{\Phi}{10^{41}{\rm
    s}^{-1}}\right)^{1/2} \left(\frac{M_{\star}}{M_{\odot}}\right)^{1/2}{\rm M_{\odot}yr^{-1}},
\end{equation}
where $\Phi$ is the stellar ionizing flux. \cite{alexander2005} found $\Phi
\sim 10^{41} - 10^{44}$ photons s$^{-1}$ in five CTTs with ages of around $10^6$
years. Assuming that the ionizing flux of T Tau N also falls within this range,
we estimate the mass loss rate to be $5\times 10^{-10} - 2 \times 10^{-8} {\rm
  M_{\odot} yr^{-1}}$.
The mass loss rate from a photoevaporative wind would thus be significantly
smaller than the accretion rates inferred from SED modeling
(Table~\ref{tab:SEDmod}), while the upper limit is similar to the observed
accretion rate of $(1.4-5.9) \times 10^{-8}$M$_{\odot}$ yr$^{-1}$
\citep{beck2004}. Since the accretion rate from the SED might be overestimated
(Sect.~\ref{geometry}) the photoevaporative
wind may have a significant effect on the dispersal of the T Tau
N disk at its present evolutionary state.

\subsection{Other indicators}
\label{indic}
Two scenarios seem likely to explain the H$_2$ emission: shocks from a
stellar wide-angle wind interacting with a flared disk, or irradiation from the
central star onto the disk. The latter may or may not be accompanied by 
photoevaporative mass loss.  
The simplest way to distinguish between shock and PDR
excitation is that the latter will create an atomic ionized layer giving rise
to hydrogen recombination lines. Furthermore, a PDR will display strong
emission from [OI] 63$\mu$m, 145$\mu$m and [CII] 158$\mu$m as well as 
[FeII] 1.26$\mu$m, 1.64$\mu$m and [OI] 6300$\AA$ and [SII]
6730$\AA$ in high density regions. We
examined the spatial distribution of the continuum-subtracted Br$\gamma$ line
to search for extended Br$\gamma$ emission in comparison to the presumably
pointlike PSF of the adjacent continuum. We did not find any
conclusive evidence for any spatial extension of the Br$\gamma$ line.   
This is consistent with \cite{kasper2002}, who found that extended
Br$\gamma$ emission, if any, is confined to 
within 6AU of the star. This argues against the existence of an extended PDR.

On the other hand, \cite{vandenancker1999} found strong 
[OI], [CII] and [FeII] lines which they were unable to fit with shock
models. Although these data suffer from low spatial resolution ($>$ 20\arcsec),
these findings suggest the presence of a PDR component.
With 2\arcsec resolution \cite{solf1999} found a compact emission region in
the optical [OI] and [SII] 
lines which is centered close to T Tau N and is characterized by near-zero
radial velocities. 
Our SINFONI observations of the T Tau system included J-band data which are not
presented here (see Gustafsson et al, in prep). These measurements are
centered on T 
Tau S and do not cover T Tau N. Nevertheless, the
data show indirect evidence of strong [FeII] 1.26$\mu$m
and [OI] 1.13$\mu$m emission close to or originating from T Tau N itself. 
In the J-band data, continuum emission from T Tau N is scattered into the
field of view and we find [OI] and [FeII]
emission with the same spatial distribution as the scattered light. We
therefore believe that the emission originates in T Tau N. The
[OI] line traces the ionization front \citep{marconi1998} and the presence of
these lines indicates that a 
PDR contribution may be present after all.

One possible solution to reconcile these apparently contradictory observations
is that a PDR exists but is confined to within $\sim$ 6AU from T Tau N. The
infrared H$_2$ emission analysed here extends to much larger radii and would
then be due to shocks. In any case, the emission seems to be linked to a nearly
face-on disk. Future observations with high spectral and spatial resolution
will be needed to confirm or disprove these suggestions. Such measurements
should 
include deeper observations of the infrared H$_2$ lines to constrain the
line ratios, spectrally resolved measurements of the velocity, a search for
extended hydrogen recombination lines, and observations 
to spatially constrain the [OI], [FeII], [SII] and [CII] lines.

\section{Implications for T Tau S}
\label{ttaus}

T Tau S is located at a projected distance of 0\farcs7 ($\sim 100 $ AU) south
of T Tau N. The extent of 
the H$_2$ emission, as well as the SED modeling, indicate that the outer radius
of the disk around T Tau N is 85-100 AU.
Disks around stars in binary or
multiple stellar systems will be truncated by their mutual gravitational
influence. The 
size of the disk depends on the mass ratio of the stars and their
separation. The disk size
typically ranges from 0.3-0.4 times the separation for mass ratios of 1-0.3,
even when the disk and orbital plane are not coplanar \citep{artymowicz1994,
  larwood1996}. If the true size of the disk around T Tau N is $\sim 100$AU,
the argumentation above
implies that the orbital distance to T Tau S is $\sim 300$AU and that the
inclination of the orbit to the line-of-sight is $\sim20^{\circ}$.   

If, on the one hand, T
Tau S is located in front of T Tau N, UV irradiation from the 2.7M$_{\odot}$
star T Tau Sa \citep{duchene2006} could 
in principle contribute significantly to the H$_2$
excitation on the side of the T Tau N disk facing toward us. 
However, in that case the H$_2$ emission would be accompanied by
  extended Br$\gamma$ emission as discussed in Sect. \ref{indic}. This is not seen.
On the other hand, if T Tau S is located behind T Tau N, the size of the disk
indicates that it 
may very well contribute to the extinction of T Tau S. 
Given the small angular separation of T Tau Sa-Sb of $\sim$0\farcs1 the disk
around T Tau N would most likely obscure both Sa and Sb by an equal amount of
material. Therefore, it does not help to explain why Sb is a normal T Tauri
star and Sa shows the characteristics of an infrared companion
\citep{dyck1982}.
The relative position of T Tau N and S and 
their outflows are the subjects of a subsequent paper (Gustafsson et al, in prep).

\section{Conclusions}
\label{conclusion}

We detect emission from the H$_2$ v=1-0 S(1) rovibrational line at 2.12$\mu$m
in a ring-like structure very close to T Tau N. 
We find that the weak H$_2$ emission is most likely linked
to a nearly face-on flared disk. Another possible solution is that the H$_2$
emission originates from shocks impacting on the lower density walls of an
envelope cavity. This scenario, however, requires that the high velocity jet
in T Tau N be less collimated than in other T Tauri stars.   
The radius of the disk is
$\sim 85-100$AU, based on SED modeling and the extent of the H$_2$ emission.
The velocity in the vicinity of T Tau N is consistent with the rest velocity
of the star to within the errors.
Both shocks associated with a wide-angle wind
impinging on the disk and UV + X-ray irradiation from the central star onto
the disk 
are plausible excitation mechanisms which can reproduce the H$_2$ flux.    
Both these mechanisms require a substantial disk around T Tau N.
However, models and observations indicate that irradiation from the central
star cannot excite H$_2$
at radii much larger than 20 AU. Thus, the most likely excitation mechanism of
H$_2$ is that of a wide-angle wind impinging on a flared disk. A PDR created by
irradiation may exist within $\sim$ 6AU from T Tau N.

\begin{acknowledgements}
We would like to thank L.E. Kristensen for providing data from numerical shock
models and J. Sauter for modeling disk and envelope densities. We are also
grateful to Reinhard Mundt and Cornelis Dullemond for fruitful discussions.

\end{acknowledgements}

\bibliographystyle{aa}
\bibliography{bibliography}

\end{document}